\documentclass{sig-alternate}
\usepackage[table]{xcolor}
\usepackage{calc}
\usepackage{float}
\usepackage{ifthen}
\usepackage{rotating}
\usepackage{graphics}
\usepackage{graphicx}
\usepackage{multirow}
\usepackage{latexsym}
\usepackage{amsmath}
\usepackage{lscape}
\usepackage{pgf}
\usepackage{tikz}
\usepackage{pgfplots}
\usetikzlibrary{trees}
\usetikzlibrary{mindmap}
\usepackage{rotating}
\usepackage{graphics}
\usepackage{graphicx}
\usepackage{multirow}
\usepackage{latexsym}
\usepackage{amsmath}
\usepackage{lscape}
\usepackage{pgf}
\usepackage{tikz}
\usepackage{qtree}
\usepackage{multirow}
\usepackage{pgfplots}
\usepackage{pgf}
\usepackage{tikz}

\begin{document}

\definecolor{light-gray}{gray}{0.90}

\newcommand{\slice}[4]{
  \pgfmathparse{0.5*#1+0.5*#2}
  \let\midangle\pgfmathresult

  \draw[thick,fill=black!10] (0,0) -- (#1:1) arc (#1:#2:1) -- cycle;

  \node[label=\midangle:#4] at (\midangle:1) {};

  \pgfmathparse{min((#2-#1-10)/110*(-0.3),0)}
  \let\temp\pgfmathresult
  \pgfmathparse{max(\temp,-0.5) + 0.8}
  \let\innerpos\pgfmathresult
  \node at (\midangle:\innerpos) {#3};
}

\title{Rhetorical Relations for Information Retrieval}
\numberofauthors{3} 
%
\author{
%
%
\alignauthor
Christina Lioma\\
       \affaddr{Computer Science}\\
       \affaddr{University of Copenhagen Denmark}\\
       \email{c.lioma@diku.dk}
\alignauthor
Birger Larsen\\
       \affaddr{Royal School of Library and Information Science Copenhagen Denmark}\\
       \email{blar@iva.dk}
\alignauthor
Wei Lu\\
       \affaddr{School of Information Management}\\
       \affaddr{Wuhan University China}\\
       \email{reedwhu@gmail.com}
}
\date{30 July 1999}

\maketitle
\begin{abstract}
Typically, every part in most coherent text has some plausible reason for its presence, some function that it performs to the overall semantics of the text. Rhetorical relations, e.g. \texttt{contrast, cause, explanation}, describe how the parts of a text are linked to each other. Knowledge about this so-called discourse structure has been applied successfully to several natural language processing tasks. This work studies the use of rhetorical relations for Information Retrieval (IR): Is there a correlation between certain rhetorical relations and retrieval performance? Can knowledge about a document's rhetorical relations be useful to IR? 

We present a language model modification that considers rhetorical relations when estimating the relevance of a document to a query. Empirical evaluation of different versions of our model on TREC settings shows that certain rhetorical relations can benefit retrieval effectiveness notably ($>10\%$ in mean average precision over a state-of-the-art baseline). 
\end{abstract}





\section{Introduction}
According to discourse analysis, every part in most coherent text tends to have some plausible reason for its presence, some function that it performs to the overall semantics of the text. Rhetorical relations, e.g. \texttt{contrast, explanation, condition}, are considered critical for text interpretation, because they signal how the parts of a text are linked to each other to form a coherent whole \cite{MannT:1988}. 
Unlike grammatical relations, which are generally explicitly manifest in language, rhetorical relations may be unstated. The goal of discourse analysis is therefore to infer rhetorical relations, and specifically to identify their span, constraints and function. 

There is a large body of research on both descriptive and predictive models of rhetorical structure and discourse analysis in natural language text. For instance, annotation projects have taken significant steps towards developing semantic \cite{FillmoreB:2002,KingsburyP:2002} and discourse \cite{CarlsonM:2003} annotated corpora. Some of these annotation efforts have already had a computational impact, making it possible to automatically induce semantic roles \cite{GildeaJ00} and to automatically identify rhetorical relations \cite{ghosh2011b}, achieving near-human levels of performance on certain tasks \cite{SoricutM03}. In addition, applications of discourse analysis to automatic language processing tasks such as summarisation or classification (overviewed in section \ref{s:relatedWork}) indicate that rhetorical relations can enhance the performance of well-trained natural language processing systems.  


\begin{figure}
\centering
\scalebox{0.20}{
\includegraphics[]{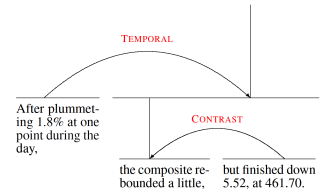}
}
\caption{\label{fig:example}Rhetorical relations example (from \cite{duVerle:2009}).}
\end{figure}

Motivated by these advances, this work brings perspectives from discourse analysis into Information Retrieval (IR) with the aim of investigating if and how rhetorical relations can benefit retrieval effectiveness. Is there a correlation between certain rhetorical relations and retrieval performance? Can knowledge about a document's rhetorical relations be useful to IR? For example, consider the rhetorical relations of the text shown in Figure \ref{fig:example} (borrowed from \cite{duVerle:2009}). Should some of the terms in this sentence be given extra weight by an IR system, according to their rhetorical relations? Can some rhetorical relations be considered more informative and hence more useful for IR ranking than others? These questions have been posed before (see discussion in section \ref{s:relatedWork}), however to our knowledge this is the first time that a principled integration of rhetorical relations into a probabilistic IR model improves precision by $>$ 10\%. 

Reasoning about query - document relevance using the language modeling formalism \cite{Croft:2003}, we present a model that conditions the probability of relevance between a query and a document on the rhetorical relations occurring in that document. We present an application of this model to an IR re-ranking task, where, given a list of documents initially retrieved for a query, the goal is to improve the ranking of the documents by refining their estimation of relevance to the query. Experimental evaluation of different versions of our model on TREC data and standard settings demonstrates that certain rhetorical relations can be beneficial to retrieval, with notable improvements to retrieval effectiveness ($>10\%$ in mean average precision and other standard TREC evaluation measures over a state-of-the-art baseline). 


\section{Related Work}
\label{s:relatedWork}

Discourse analysis and rhetorical structures have been studied in the context of several automatic text processing applications. This has been partly enabled by the availability of discourse parsers - see \cite{duVerle:2009,ghosh2011b} for up-to-date overviews of discourse parsing technology. Studies of discourse analysis in relation to IR and its broader applications are briefly overviewed below. For a more general overview of discourse analysis approaches, see Wang et al. \cite{WangLKNB11}, section 2.

   Sun \& Chai \cite{SunC:2007} investigate the role of discourse processing and its implication on query expansion for a sequence of questions in scenario-based context question answering (QA). They consider a sequence of questions as a mini discourse. An empirical examination of three discourse theoretic models indicates that their discourse-based approach can significantly improve QA performance over a baseline of plain reference resolution. 

In a different task, Wang et al. \cite{WangLKNB11} parse Web user forum threads to determine the discourse dependencies between posts in order to improve information access over Web forum archives. They present three different methods for classifying the discourse relationships between posts, which are found to outperform an informed baseline.

Heerschop et al. \cite{HeerschopG:2011} perform document sentiment analysis (partly) based on a document's discourse structure. They hypothesise that by splitting a text into important and less important text spans, and by subsequently making use of this information by weighting the sentiment conveyed by distinct text spans in accordance with their importance, they can improve the performance of a sentiment classifier. A document's discourse structure is obtained by applying rhetorical structure theory on a sentence level. They report a 4.5\% improvement in sentiment classification accuracy when considering discourse, in comparison to a non-discourse based baseline. Similarly to this study, Somasundaran et al. \cite{Som09} report improvements to opinion polarity classification when using discourse, and Morato et al. \cite{MoratoL:2003} 
report a positive dependence between classification performance and certain discourse variables. An overview of discourse analysis for opinion detection can be found in Zhou et al. \cite{ZhouLGWW11}.

In the area of text compression, Louis et al. \cite{LouisJN10} study the usefulness of rhetorical relations between sentences for summarisation. They find that most of the significant rhetorical relations are associated to non-discriminative sentences, i.e. sentences that are \textit{not important} for summarisation. They report that rhetorical relations that may be intuitively perceived as highly salient do not provide strong indicators of informativeness; instead, the usefulness of rhetorical relations is in providing constraints for navigating through the text's structure. These findings are compatible with the study of Clarke \& Lapata \cite{ClarkeL:2010} into constraining text compression on the basis of rhetorical relations. For a more in-depth look into the impact of individual rhetorical relations to summarisation see Teufel \& Moens \cite{TeufelM02}. 

In domain-specific IR, Yu et al. \cite{YuW:2009} focus on psychiatric document retrieval, which aims to assist users to locate documents relevant to their depressive problems. They propose the use of high-level discourse information extracted from queries and documents, such as negative life events, depressive symptoms and semantic relations between symptoms, to improve the precision of retrieval results. Their discourse-aware retrieval model achieves higher precision than the vector space and Okapi models.

Closer to our work, Wang et al. \cite{WangLWK06} extend an IR ranking model by adding a re-ranking strategy based on document discourse. Specifically, their re-ranking formula consists of the original retrieval status value computed with the BM11 model, which is then multiplied by a function that linearly combines inverse document frequency and term distance for each query term within a discourse unit. They focus on one discourse type only (\texttt{advantage-disadvantage}) which they identify manually in queries, and show that their approach improves retrieval performance for these queries. Our work differs on several points. We use an automatic (not manual) discourse parser to identify rhetorical relations in the documents to be retrieved (not queries). We consider 15 rhetorical relations (not 1) and we study their impact to retrieval performance using a modification of the IR language model.

Finally, Suwandaratna \& Perera \cite{SuP:2010} also present a re-ranking approach for Web search that uses discourse structure. They report a heuristic algorithm for refining search results based on their rhetorical relations. Their implementation and evaluation is partly based on a series of ad-hoc choices, making it hard to compare with other approaches. They report a positive user-based evaluation of their system for ten test cases. 


\section{Ranking with Rhetorical \\ Relations}
\label{s:Model}
There may be various ways of considering rhetorical relations in an IR setting. In this work, we view rhetorical relations as non-overlapping text spans, rather than a graph or a tree with structure and overlapping nodes \cite{SoricutM03}. We select a principled integration of rhetorical relation information into the retrieval model that ranks documents with respect to queries. The goal is to enable evidence about the rhetorical relations in a document to have a quantifiable impact upon the estimation of relevance of this document to a query, and to study that impact.

\subsection{Model Derivation}
\label{ss:Derivation}
Let $q$ be a query, $d$ a document, $D$ a collection of documents, and $\psi_g$ a rhetorical relation in the collection (so that $\sum_{\psi_g} p(\psi_g|d)=1$). In probabilistic IR, each $d$ in $D$ can be ranked by its probability $p(d|q)$ of being relevant to $q$. Using Bayes' law:
\begin{equation}
\label{eq:1}
p(d|q) = \frac{p(q|d)p(d)}{p(q)} \stackrel{rank}{=} p(q|d)
\end{equation}
\noindent where the right-hand side of Equation \ref{eq:1} is derived as follows: $p(q)$ is dropped because it is fixed for all documents, and $p(d)$ can be dropped on the assumption that it is uniform in the absence of any prior knowledge about any document. Using the language modeling approach to IR \cite{Croft:2003}, $p(q|d)$ can be interpreted as the probability of generating the terms in $q$ from a model induced by $d$, or more simply how likely it is that the document is about the same topic as the query. $p(q|d)$ can be estimated in different ways, for instance using Dirichlet, Jelinek-Mercer, or two-stage smoothing \cite{ZhaiL02}. 

We introduce into Equation \ref{eq:1} the probability of generating the query terms from a model induced by $d$ and by its rhetorical relations $\psi \in d$ as follows:

\begin{equation}
\label{eq:2}
p(q|d) = \sum_{\psi_g} p(q|d,\psi_g) p(\psi_g|d) 
\end{equation}

We now explain the two components in Equation \ref{eq:2}. The first component, $p(q|d,\psi_g)$, can be interpreted as the probability of generating the query terms from a model induced by $d$ and $\psi_g$. We estimate $p(q|d,\psi_g)$ as a simple mixture of the probabilities of generating $q$ from $d$ and $\psi_g$: 
\begin{equation}
\label{eq:mix}
p(q|d,\psi_g) = (1 - \kappa) \cdot  p(q|d) + \kappa \cdot  p(q|\psi_g)
\end{equation} 

\noindent where $p(q|d)$ is the (baseline) probability of relevance between $q$ and $d$ mentioned in the beginning of this section, $\kappa$ is a free parameter, and $p(q|\psi_g)$ can be interpreted as the probability of generating $q$ from a model induced by the rhetorical relation $\psi_g$, or more simply, the `likelihood of relevance' between the terms in the query and the terms in the rhetorical relation. 

The second component of Equation \ref{eq:2}, $p(\psi_g|d)$, is the probability of the rhetorical relation given the document. Similarly to above, this can be interpreted as the probability of generating the terms in $\psi_g$ from a model induced by $d$, or more simply the likelihood of relevance between the terms in the rhetorical relation and the terms in the document.

\begin{table*}
\centering
\caption{\label{tab:rr}Examples of the 15 rhetorical relations (in bold italics) of our dataset, identified by the SPADE discourse parser \cite{SoricutM03}}
\begin{tabular}{|l|l|} 
\hline
Rhetorical relation&Example sentences with rhetorical relations italicised and bold \\ 
\hline
attribution&... the islands now known as the Gilbert Islands were settled \textbf{\textit{by Austronesian-speaking people}} ... \\
background&... many whites had left the country \textbf{\textit{when Kenyatta divided their land among blacks}} ... \\
cause-result&... \textbf{\textit{I plugged ``wives'' into the search box and came up with the following results}} ... \\
comparison&... so for humans, \textbf{\textit{it is stronger than coloured}} to frustrate these unexpected numbers ... \\
condition&... Conditional money \textbf{\textit{based upon care for the pet}} ... \\
consequence&... voltage drop with the cruise control switch \textbf{\textit{could cause erratic cruise control operation}} ... \\
contrast&... \textbf{\textit{Although it started out as a research project}}, the ARPANET quickly developed into ...\\
elaboration&... order accutane \textbf{\textit{no prescription required}} ... \\
enablement&... The project will also \textbf{\textit{offer exercise programs and make eye care services accessible}} ... \\
evaluation&... such advances will be reflected in an ever-\textbf{\textit{greater proportion of grade A recommendations}} ... \\
explanation&... the concept \textbf{\textit{called as ``evolutionary developmental biology'' or shortly ``evo-devo''}} ... \\
manner-means&... Fill current path \textbf{\textit{using even-odd rule, then paint the path}} ... \\
summary&... Safety Last, Girl Shy, Hot Water, The Kid Brother, Speedy \textbf{\textit{(all with lively orchestral scores)}} ... \\
temporal&... Take time out \textbf{\textit{before you start writing}} ...\\
topic-comment&... \textbf{\textit{Director Mark Smith expressed support for greyhound adoption}} ... \\
\hline
\end{tabular}
\end{table*}

\subsection{Model Induction}
\label{ss:Induction}
To make Equations \ref{eq:2}-\ref{eq:mix} operational we need to compute $p(q|\psi)$ and $p(\psi|d)$. One simple way of doing so is using the respective maximum likelihood estimations:

\begin{equation}
\label{eq:3}
\log p(q|\psi_g) = \sum^{|q|}_{i=1}\frac{f(q_i,\psi_g)}{|\psi_g|}
\end{equation}
\noindent where $f(q_i,\psi_g)$ is the frequency of the query term $q_i$ in $\psi_g$, and $|\psi_g|$ is the number of terms in $\psi_g$. 

\begin{equation}
\label{eq:4}
\log p (\psi_g|d) = \sum^{|\psi_g|}_{j=1} \frac{f(\psi_{gj},d)}{|d|}
\end{equation}
\noindent where $f(\psi_{gj},d)$ is the frequency of the rhetorical relation term $\psi_{gj}$ in $d$, and $|d|$ is the number of terms in $d$.  In this work, we use the above equations and, to compensate for zero-frequency cases, we apply add-one smoothing. 

Alternative principled estimations of Equations \ref{eq:3}-\ref{eq:4} are possible (e.g. Dirichlet, Good-Turing) and could potentially improve the performance reported in this work. For instance, one could discount the frequencies in Equations \ref{eq:3}-\ref{eq:4} by a respective collection model using Dirichlet smoothing: $\log p_s(q|\psi_g) = \sum^{|q|}_{i=1} \frac{f(q_i,\psi_g) + \mu \cdot p(q_i|\Psi)}{|\psi_g|+\mu}$ where $\mu$ would be the smoothing parameter and $\Psi$ would be the collection of all rhetorical relations in $D$. A similarly Dirichlet smoothed alternative estimation of Equation \ref{eq:4} would be: $\log p_s(\psi_g|d) = \sum^{|\psi_g|}_{j=1} \frac{f(\psi_{gj},d) + \mu \cdot p(\psi_{gj}|D)}{|d|+\mu}$. We choose to use maximum likelihood instead of Dirichlet to avoid introducing the extra Dirichlet smoothing parameter $\mu$ when investigating the effect of rhetorical relations upon retrieval. 

Another alternative would be to use Good-Turing smoothing, however doing so would scale down the maximum likelihood estimations in Equations \ref{eq:3}-\ref{eq:4} by a factor of $1 - \frac{E(1)}{|\psi_g|}$ and $1 - \frac{E(1)}{|d|}$ respectively, where $\frac{E(1)}{|\psi_g|}$ (resp. $\frac{E(1)}{|d|}$) is the estimate of how many items in the numerator of Equation \ref{eq:3} (resp. Equation \ref{eq:4}) have occurred once in the sample of the denominator (see Gale \& Sampson \cite{GaleS95} for more on Good-Turing smoothing). In effect, for Equation \ref{eq:3} this scaling down would reduce the probability of the query terms that we have seen in $\psi_g$, making room for query terms that we have not seen. For our setting this would not be necessary, because in practice most queries and most rhetorical relations correspond to rather short text spans. Good-Turing smoothing might be better suited for larger samples \cite{GaleS95}.
 
Overall, the model presented in this section can be seen as a `basic model' for ranking documents (partly) according to their rhetorical relations. Different variations on this basic model are certainly possible, however we choose to use the simple maximum likelihood version of this model for this exploratory investigation into the potential benefits of using rhetorical relations for IR.


\section{Evaluation}
\label{s:Evaluation}

\subsection{Experimental Setup}
\label{ss:Settings}

We evaluate our model on the task of re-ranking an initial list of documents, which has been retrieved in response to a query. Re-ranking is a well-known IR practice that can enhance retrieval performance notably \cite{KrikonK11}. The baseline of our experiments consists of the top 1000 documents retrieved for each query using a state-of-the-art retrieval model (language model with Dirichlet smoothing\footnote{We also experimented with Jelinek-Mercer and two-stage smoothing for the baseline retrieval model. Dirichlet and two-stage gave higher scores. We chose Dirichlet over two-stage because it includes one less parameter to tune.} \cite{Croft:2003}). Our approach reranks these documents using Equation \ref{eq:2}.

\subsubsection{Dataset and Pre-processing}
We experiment with the TREC datasets of the Web 2009 (queries 1-50) and Web 2010 (queries 51-100) tracks, that contain collectively 100 queries and their relevance assessments on the Clueweb09 cat. B dataset\footnote{http://lemurproject.org/clueweb09.php/} (50,220,423 web pages in English crawled between January and February 2009). We choose these datasets because they are used widely in the community, allowing comparisons with state-of-the-art. We remove spam using the spam rankings of Cormack et al. \cite{cormackS:2010} with the recommended setting of percentile-score $<70$ indicating spam\footnote{Note that removing spam from Clueweb09 cat B. is known to give overall lower retrieval scores than keeping spam \cite{BenderskyCD11}.}. 

We consider a subset of this collection, consisting of the top 1000 documents that have been retrieved in response to each query by the baseline retrieval model on tuned settings (described in section \ref{sss:tuning}) using the Indri IR system\footnote{http://www.lemurproject.org/} for indexing and retrieval. 
For this subset, we strip HTML annotation using our in-house WHU-REAPER crawling and web parsing toolkit\footnote{Freely available by emailing the third author.}. Rhetorical relations are identified using the freely available SPADE discourse parser \cite{SoricutM03}. Table \ref{tab:rr} shows the 15 types of rhetorical relations identified by this process, with examples taken from the re-ranking dataset. 

\subsubsection{Parameter Tuning}
\label{sss:tuning}
Two parameters are involved in these experiments: the Dirichlet smoothing parameter $\mu$ of the retrieval model (used by both the baseline and our approach) and the mixture parameter $\kappa$ of our model. Both parameters are tuned using 5-fold cross validation for each query set separately; results reported are the average over the five test sets. $\mu$ is tuned across \{100, 500, 800, 1000, 2000, 3000, 4000, 5000, 8000, 10000\} (using the range of Zhai \& Lafferty \cite{ZhaiL02}) and $\kappa$ is tuned across \{0.1, 0.3, 0.5, 0.7, 0.9\}. 

Performance is reported and tuned separately for Mean Average Precision (MAP), Binary Preference (BPREF), and Normalised Discounted Cumulated Gain (NDCG). These measures contribute different aspects to the overall evaluation: BPREF measures the average precision of a ranked list; it differs from MAP in that it does not treat non-assessed documents as explicitly non-relevant (whereas MAP does) \cite{BuckleyV04}. This is a useful insight, especially for a collection as large as Clueweb09 cat. B where the chances of retrieving non-assessed documents are higher. NDCG measures the gain of a document based on its position in the result list. The gain is accumulated from the top of the ranked list to the bottom, with the gain of each document discounted at lower ranks. This gain is relative to the ideal based on a known recall base of relevance assessments \cite{JarvelinK02}. Finally, we test the statistical significance of our results using the t-test at 95\% and 99\% confidence levels \cite{SmuckerAC09}.




\begin{table*}
\centering
\caption{\label{tab:scores}Retrieval performance with rhetorical relations and without (baseline). * (**) marks stat. significance at 95\% (99\%) using the t-test. Bold means $>$ baseline. \% shows the difference from the baseline. Shaded rows indicate consistent improvements over the baseline at all times.}
\scalebox{0.8}{
\begin{tabular}{|l||lr|lr|lr||lr|lr|lr|} 
\hline
\multirow{2}{*}{rhetorical relation} 
&\multicolumn{6}{c||}{Web 2009 (queries 1-50)} &\multicolumn{6}{c|}{Web 2010 (queries 51-100)}\\
    &\multicolumn{2}{c|}{MAP}    	&\multicolumn{2}{c|}{BPREF}  	&\multicolumn{2}{c||}{NDCG}  &\multicolumn{2}{c|}{MAP}    	&\multicolumn{2}{c|}{BPREF}  	&\multicolumn{2}{c|}{NDCG}  \\ 
\hline
none (baseline) &0.1625&  	        &0.3230& 	            &0.3893&            &0.0986&            &0.2240&             &0.2920&\\
attribution 	&\bf0.1654*&+1.8\% 	&\bf0.3275**&+1.4\% 	&\bf0.3927**&+0.9\% &0.0924&-6.2\%      &\bf0.2549**&+13.8\% &\bf0.3008**&+3.0\%\\
\rowcolor{light-gray}
background 	    &\bf0.1646&+1.3\% 	&\bf0.3291**&+1.9\%	    &\bf0.3910&+0.4\%   &\bf0.1086*&+10.2\% &\bf0.2623**&+17.1\% &\bf0.3070**&+5.1\%\\
\rowcolor{light-gray}
cause-result	&\bf0.1626&+0.1\% 	&\bf0.3255**&+0.8\% 	&\bf0.3900&+0.2\%   &\bf0.1015&+2.9\%   &\bf0.2491*&+11.2\%  &\bf0.3079&+5.4\%\\
comparison 	    &0.1610&-0.9\%		&\bf0.3251*&+0.6\% 	    &0.3877&-0.4\%      &\bf0.1017&+3.1\%   &\bf0.2282&+1.9\%    &\bf0.3040**&+4.1\%\\
\rowcolor{light-gray}
condition 	    &\bf0.1632&+0.5\% 	&\bf0.3258**&+0.9\% 	&\bf0.3903&+0.3\%   &\bf0.0999&+1.3\%   &\bf0.2470**&+10.3\% &\bf0.2936&+0.5\%\\
consequence 	&0.1602&-1.4\% 	    &\bf0.3250&+0.6\% 	    &0.3874&-0.5\%      &0.0945&-4.1\%      &\bf0.2377*&+6.1\%   &\bf0.2840**&-2.7\%\\
contrast 	    &0.1549*&-4.6\% 	&\bf0.3269**&+1.2\% 	&\bf0.3897&+0.1\%   &\bf0.1103*&+11.8\% &\bf0.2531**&+13.0\% &\bf0.3069**&+5.1\%\\
elaboration 	&0.1556*&-4.2\% 	&\bf0.3292**&+1.9\% 	&0.3866&-0.7\%      &0.0951&-3.5\%      &\bf0.2598**&+16.0\% &\bf0.3005**&+2.9\%\\
enablement 	    &0.1601&-1.4\% 	    &\bf0.3240&+0.3\% 	    &0.3869*&-0.6\%     &\bf0.1010&+2.4\%   &\bf0.2316*&+3.4\%   &\bf0.2992*&+2.5\%\\
evaluation 	    &\bf0.1632&+0.5\% 	&\bf0.3242&+0.4\% 	    &0.3886&-0.2\%      &0.0814**&-17.4\%   &\bf0.2313*&+3.3\%   &\bf0.2902&-0.6\%\\
explanation 	&0.1546&-4.9\% 	    &\bf0.3259*&+0.9\% 	    &0.3813&-2.1\%      &\bf0.1034&+4.9\%   &\bf0.2645**&+18.1\% &\bf0.3069**&+5.1\%\\
manner-means 	&0.1623&-0.1\% 	    &\bf0.3253*&+0.7\% 	    &0.3884&-0.2\%      &0.0986&-           &\bf0.2324*&+3.7\%   &\bf0.2897&-0.8\%\\
summary 	    &\bf0.1626&+0.1\% 	&\bf0.3241&+0.3\% 	    &0.3879&-0.4\%      &0.0862&-12.6\%     &0.2220*&-0.9\%      &\bf0.2928&+0.3\%\\
temporal	    &0.1615&-0.6\% 	    &\bf0.3262**&+1.0\% 	&0.3887&-0.2\%      &0.0921&-6.6\%      &\bf0.2546**&+13.7\% &\bf0.3052&+4.5\%\\
\rowcolor{light-gray}
topic-comment 	&\bf0.1673&+3.0\% 	&\bf0.3375&+4.5\% 	    &\bf0.3976*&+2.1\%  &\bf0.1090*&+10.5\% &\bf0.2476*&+10.5\%  &\bf0.3009&+3.1\%\\
\hline
\end{tabular}
}
\end{table*}

\subsection{Findings}
\label{ss:Results}

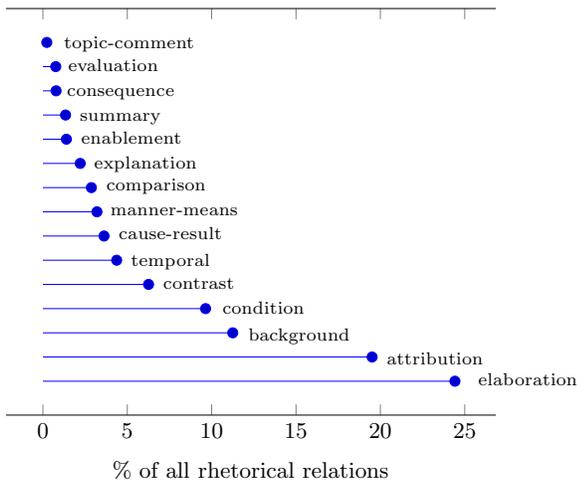
\begin{figure}
\centering
\scalebox{0.95}{
\begin{tikzpicture}
\begin{axis}[
yticklabels= ,
hide y axis=true,
xlabel=  \% of all rhetorical relations
]
\addplot+[xcomb] coordinates
{
(24.42,0)
(19.51,1) 
(11.25,2) 
(9.64,3)
(6.26,4)
(4.37,5)
(3.62,6)
(3.2,7)
(2.87,8)
(2.21,9)
(1.39,10)
(1.34,11)
(0.78,12)
(0.76,13)
(0.23,14)
};
\end{axis}
\draw (7.3,0.5) node[] { \scriptsize elaboration};
\draw (6.0,0.8) node[] { \scriptsize attribution};
\draw (4.1,1.1) node[] { \scriptsize background};
\draw (3.6,1.5) node[] { \scriptsize condition};
\draw (2.7,1.85) node[] { \scriptsize contrast};
\draw (2.3,2.15) node[] { \scriptsize temporal};
\draw (2.3,2.55) node[] { \scriptsize cause-result};
\draw (2.35,2.85) node[] { \scriptsize manner-means};
\draw (2.1,3.2) node[] { \scriptsize comparison};
\draw (1.95,3.52) node[] { \scriptsize explanation};
\draw (1.75,3.88) node[] { \scriptsize enablement};
\draw (1.6,4.15) node[] { \scriptsize summary};
\draw (1.6,4.5) node[] { \scriptsize consequence};
\draw (1.5,4.9) node[] { \scriptsize evaluation};
\draw (1.72,5.2) node[] { \scriptsize topic-comment};
 \end{tikzpicture}
}
\caption{\label{fig:distr-rr} \% distribution of rhetorical relations in our dataset.}
\end{figure}

Figure \ref{fig:distr-rr} shows the distribution of the rhetorical relations in our re-ranking dataset as a percentage of the total number of rhetorical relations. \texttt{Elaboration, attribution} and \texttt{background} are the most frequent rhetorical relations, whereas \texttt{topic-comment} is the most infrequent. This happens because quite often in text a topic forms the nucleus of the discourse, which is then linked by a number of different rhetorical relations, for instance about its background, elaborating on an aspect, or attributing parts of it to some entity. As a result, several types of other rhetorical relations can correspond to a single \texttt{topic-comment}.  Note that the distribution of rhetorical relations reported here is in agreement with the literature, e.g. Teufel \& Moens \cite{TeufelM02} also report a 5\% occurrence of \texttt{contrast}, albeit in the domain of scientific articles.        

\subsubsection{Retrieval-Enhancing Rhetorical Relations}
Table \ref{tab:scores} shows the performance of our model against the baseline, for each rhetorical relation and evaluation measure. The baseline performance is among the highest reported in the literature for these setings; for instance Bendersky et al. \cite{BenderskyCD11} report MAP=0.1605 for a tuned language model baseline with the Web 2009 track queries on Clueweb cat. B without spam. 

We observe that different rhetorical relations perform differently across evaluation measures and query sets. The four rhetorical relations that improve performance over the baseline consistently for all evaluation measures and query sets (shaded rows in Table \ref{tab:scores}) are: \texttt{background, cause-result, condition} and \texttt{topic-comment}. \texttt{Topic-comment} is one of the overall best-performing rhetorical relations, which in simple terms means that boosting the weight of the topical part of a document improves its estimation of relevance. 

\begin{table}
\centering
\caption{\label{tab:kappa} Effect of the rhetorical relation to the retrieval model as indicated by parameter $\kappa$ (see Equation \ref{eq:mix}), for the tuned runs of Table \ref{tab:scores}. Shaded rows indicate rhetorical relations that consistently improve performance over the baseline at all times.}
\scalebox{0.77}{
\begin{tabular}{|l||l|l|l||l|l|l|} 
\hline
rhetorical &\multicolumn{3}{c||}{Web 2009 (queries 1-50)} &\multicolumn{3}{c|}{Web 2010 (queries 51-100)}\\
relation    &MAP    	&BPREF  	&NDCG &MAP    	&BPREF  	&NDCG  \\ 
\hline
attribution 	&0.1   &0.5   &0.1  &0.3    &0.5   &0.3\\
\rowcolor{light-gray}
background 	    &0.2   &0.6   &0.2  &0.3    &0.7   &0.3\\
\rowcolor{light-gray}
cause-result	&0.3   &0.7   &0.3  &0.5    &0.7   &0.5\\
comparison 	    &0.4   &0.7   &0.4  &0.3    &0.5   &0.3\\
\rowcolor{light-gray}
condition 	    &0.3   &0.7   &0.3  &0.3    &0.5   &0.3\\
consequence 	&0.5   &0.7   &0.5  &0.5    &0.7   &0.5\\
contrast 	    &0.3   &0.7   &0.3  &0.3    &0.5   &0.3\\
elaboration 	&0.1   &0.5   &0.1  & 0.3   &0.5   &0.3\\
enablement 	    &0.1   &0.9   &0.1  &0.3    &0.5   &0.3\\
evaluation 	    &0.5   &0.7   &0.5  &0.5    &0.7   &0.5\\
explanation 	&0.5   &0.7   &0.5  &0.5    &0.7   &0.5\\
manner-means 	&0.5   &0.7   &0.5  &0.5    &0.7   &0.5\\
summary 	    &0.5   &0.7   &0.5  &0.3    &0.7   &0.3\\
temporal	    &0.1   &0.7   &0.1  &0.3    &0.5   &0.3\\
\rowcolor{light-gray}
topic-comment 	&0.5   &0.5  &0.5  &0.5    &0.7   &0.5\\
\hline
\end{tabular}
}
\end{table}

A closer look at which rhetorical relations decrease performance presents a more uneven picture as no relations consistently underperform for all measures and query sets. Some relations, such as \texttt{explanation} and \texttt{enablement} for Web 2009, and \texttt{summary} and \texttt{evaluation} for Web 2010, are among the lowest performing, but are not under the baseline across all measures and both query sets. This implies that separating rhetorical relations into those that generally can enhance retrieval performance and those that cannot may not be straight-forward. Even though exploring the family likeness between useful relations and ones that give no mileage is an interesting discussion, in the rest of the paper we focus on those rhetorical relations that consistently improve retrieval performance (for these datasets).

Improvements over the baseline are generally higher for Web 2010 than Web 2009, possibly because the former baseline is weaker, with potentially more room for improvement. An interesting trend is that more rhetorical relations improve performance according to BPREF than according to MAP and NDCG. As BPREF is the only of these evaluation measures that does not consider non-assessed documents as non-relevant, this indicates the presence of non-assessed documents in the ranking. 

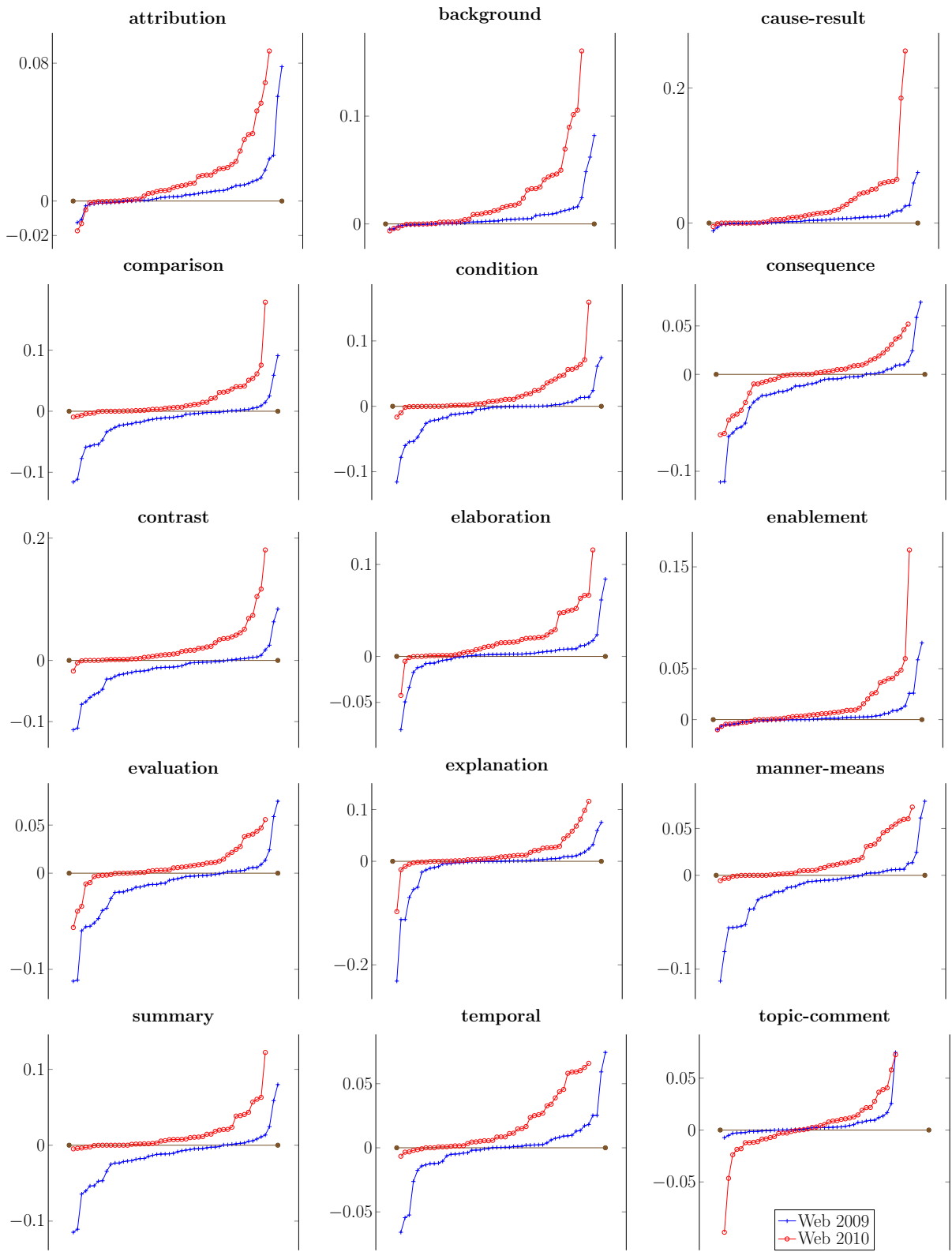
\begin{figure*}
\centering
\pgfplotsset{every axis label/.append style={font=\tiny}}
\tikzset{every mark/.append style={font=\tiny}}
\pgfplotsset{every axis legend/.append style={
			at={(0.5,-0.0)},
			anchor=south}
			}
\scalebox{1.0}{
\begin{tabular}{ccc}
\begin{tikzpicture}[baseline,scale=0.5]
\begin{axis}[
scale only axis,
ytick={-0.02,0.0,+0.08},
hide x axis=true,
font=\LARGE,
scaled y ticks = false,
y tick label style={/pgf/number format/fixed},
title=\textbf{attribution}
]
\pgfplotstableread{plot-diff-attribution-base-map}\table 
\addplot+[mark=+] table[x index=0,y index=1] from \table;
\pgfplotstableread{plot-diff-2010-attribution-base-map}\table 
\addplot+[mark=o] table[x index=0,y index=1] from \table;
\addplot+[smooth] coordinates{(0,0) (50,0)};
\end{axis}
\end{tikzpicture} &
\begin{tikzpicture}[baseline,scale=0.5]
\begin{axis}[
ytick={0.0,+0.1},
scale only axis,
hide x axis=true,
font=\LARGE,
scaled y ticks = false,
y tick label style={/pgf/number format/fixed},
title=\textbf{background}
]
\pgfplotstableread{plot-diff-background-base-map}\table 
\addplot+[mark=+] table[x index=0,y index=1] from \table;
\pgfplotstableread{plot-diff-2010-background-base-map}\table 
\addplot+[mark=o] table[x index=0,y index=1] from \table;
\addplot+[smooth] coordinates{(0,0) (50,0)};
\end{axis}
\end{tikzpicture} &
\begin{tikzpicture}[baseline,scale=0.5]
\begin{axis}[
ytick={0.0,+0.2},
scale only axis,
hide x axis=true,
scaled y ticks = false,
y tick label style={/pgf/number format/fixed},
font=\LARGE,
title=\textbf{cause-result}
]
\pgfplotstableread{plot-diff-cause-base-map}\table 
\addplot+[mark=+] table[x index=0,y index=1] from \table;
\pgfplotstableread{plot-diff-2010-cause-base-map}\table 
\addplot+[mark=o] table[x index=0,y index=1] from \table;
\addplot+[smooth] coordinates{(0,0) (50,0)};
\end{axis}
\end{tikzpicture} \\
\begin{tikzpicture}[baseline,scale=0.5]
\begin{axis}[
ytick={-0.1,0.0,+0.1},
scale only axis,
hide x axis=true,
scaled y ticks = false,
y tick label style={/pgf/number format/fixed},
font=\LARGE,
title=\textbf{comparison}
]
\pgfplotstableread{plot-diff-comparison-base-map}\table 
\addplot+[mark=+] table[x index=0,y index=1] from \table;
\pgfplotstableread{plot-diff-2010-comparison-base-map}\table 
\addplot+[mark=o] table[x index=0,y index=1] from \table;
\addplot+[smooth] coordinates{(0,0) (50,0)};
\end{axis}
\end{tikzpicture}&
\begin{tikzpicture}[baseline,scale=0.5]
\begin{axis}[
ytick={-0.1,0,0.1},
ytick={-0.1,0.0,+0.1},
scale only axis,
hide x axis=true,
scaled y ticks = false,
y tick label style={/pgf/number format/fixed},
font=\LARGE,
title=\textbf{condition}
]
\pgfplotstableread{plot-diff-condition-base-map}\table 
\addplot+[mark=+] table[x index=0,y index=1] from \table;
\pgfplotstableread{plot-diff-2010-condition-base-map}\table 
\addplot+[mark=o] table[x index=0,y index=1] from \table;
\addplot+[smooth] coordinates{(0,0) (50,0)};
\end{axis}
\end{tikzpicture} &
\begin{tikzpicture}[baseline,scale=0.5]
\begin{axis}[
ytick={-0.1,0.0,+0.05},
scale only axis,
hide x axis=true,
scaled y ticks = false,
y tick label style={/pgf/number format/fixed},
font=\LARGE,
title=\textbf{consequence}
]
\pgfplotstableread{plot-diff-consequence-base-map}\table 
\addplot+[mark=+] table[x index=0,y index=1] from \table;
\pgfplotstableread{plot-diff-2010-consequence-base-map}\table 
\addplot+[mark=o] table[x index=0,y index=1] from \table;
\addplot+[smooth] coordinates{(0,0) (50,0)};
\end{axis}
\end{tikzpicture} \\
\begin{tikzpicture}[baseline,scale=0.5]
\begin{axis}[
ytick={-0.1,0.0,+0.2},
scale only axis,
hide x axis=true,
scaled y ticks = false,
y tick label style={/pgf/number format/fixed},
font=\LARGE,
title=\textbf{contrast}
]
\pgfplotstableread{plot-diff-contrast-base-map}\table 
\addplot+[mark=+] table[x index=0,y index=1] from \table;
\pgfplotstableread{plot-diff-2010-contrast-base-map}\table 
\addplot+[mark=o] table[x index=0,y index=1] from \table;
\addplot+[smooth] coordinates{(0,0) (50,0)};
\end{axis}
\end{tikzpicture} &
\begin{tikzpicture}[baseline,scale=0.5]
\begin{axis}[
ytick={-0.05,0.0,+0.1},
scale only axis,
hide x axis=true,
font=\LARGE,
scaled y ticks = false,
y tick label style={/pgf/number format/fixed},
title=\textbf{elaboration}
]
\pgfplotstableread{plot-diff-elaboration-base-map}\table 
\addplot+[mark=+] table[x index=0,y index=1] from \table;
\pgfplotstableread{plot-diff-2010-elaboration-base-map}\table 
\addplot+[mark=o] table[x index=0,y index=1] from \table;
\addplot+[smooth] coordinates{(0,0) (50,0)};
\end{axis}
\end{tikzpicture} &
\begin{tikzpicture}[baseline,scale=0.5]
\begin{axis}[
ytick={0.0,+0.05,0.15},
scale only axis,
hide x axis=true,
scaled y ticks = false,
y tick label style={/pgf/number format/fixed},
font=\LARGE,
title=\textbf{enablement}
]
\pgfplotstableread{plot-diff-enablement-base-map}\table 
\addplot+[mark=+] table[x index=0,y index=1] from \table;
\pgfplotstableread{plot-diff-2010-enablement-base-map}\table 
\addplot+[mark=o] table[x index=0,y index=1] from \table;
\addplot+[smooth] coordinates{(0,0) (50,0)};
\end{axis}
\end{tikzpicture} \\
\begin{tikzpicture}[baseline,scale=0.5]
\begin{axis}[
ytick={-0.1,0.0,+0.05},
scale only axis,
hide x axis=true,
font=\LARGE,
scaled y ticks = false,
y tick label style={/pgf/number format/fixed},
title=\textbf{evaluation}
]
\pgfplotstableread{plot-diff-evaluation-base-map}\table 
\addplot+[mark=+] table[x index=0,y index=1] from \table;
\pgfplotstableread{plot-diff-2010-evaluation-base-map}\table 
\addplot+[mark=o] table[x index=0,y index=1] from \table;
\addplot+[smooth] coordinates{(0,0) (50,0)};
\end{axis}
\end{tikzpicture} &
\begin{tikzpicture}[baseline,scale=0.5]
\begin{axis}[
ytick={-0.2,0.0,+0.1},
scale only axis,
hide x axis=true,
font=\LARGE,
scaled y ticks = false,
y tick label style={/pgf/number format/fixed},
title=\textbf{explanation}
]
\pgfplotstableread{plot-diff-explanation-base-map}\table 
\addplot+[mark=+] table[x index=0,y index=1] from \table;
\pgfplotstableread{plot-diff-2010-explanation-base-map}\table 
\addplot+[mark=o] table[x index=0,y index=1] from \table;
\addplot+[smooth] coordinates{(0,0) (50,0)};
\end{axis}
\end{tikzpicture} &
\begin{tikzpicture}[baseline,scale=0.5]
\begin{axis}[
ytick={-0.1,0.0,+0.05},
scale only axis,
hide x axis=true,
font=\LARGE,
scaled y ticks = false,
y tick label style={/pgf/number format/fixed},
title=\textbf{manner-means}
]
\pgfplotstableread{plot-diff-mannerxmeans-base-map}\table 
\addplot+[mark=+] table[x index=0,y index=1] from \table;
\pgfplotstableread{plot-diff-2010-mannerxmeans-base-map}\table 
\addplot+[mark=o] table[x index=0,y index=1] from \table;
\addplot+[smooth] coordinates{(0,0) (50,0)};
\end{axis}
\end{tikzpicture} \\
\begin{tikzpicture}[baseline,scale=0.5]
\begin{axis}[
ytick={-0.1,0.0,+0.1},
scale only axis,
hide x axis=true,
font=\LARGE,
scaled y ticks = false,
y tick label style={/pgf/number format/fixed},
title=\textbf{summary}
]
\pgfplotstableread{plot-diff-summary-base-map}\table 
\addplot+[mark=+] table[x index=0,y index=1] from \table;
\pgfplotstableread{plot-diff-2010-summary-base-map}\table 
\addplot+[mark=o] table[x index=0,y index=1] from \table;
\addplot+[smooth] coordinates{(0,0) (50,0)};
\end{axis}
\end{tikzpicture} &
\begin{tikzpicture}[baseline,scale=0.5]
\begin{axis}[
ytick={-0.05,0.0,+0.05},
scale only axis,
hide x axis=true,
font=\LARGE,
scaled y ticks = false,
y tick label style={/pgf/number format/fixed},
title=\textbf{temporal}
]
\pgfplotstableread{plot-diff-temporal-base-map}\table 
\addplot+[mark=+] table[x index=0,y index=1] from \table;
\pgfplotstableread{plot-diff-2010-temporal-base-map}\table 
\addplot+[mark=o] table[x index=0,y index=1] from \table;
\addplot+[smooth] coordinates{(0,0) (50,0)};
\end{axis}
\end{tikzpicture} &
\begin{tikzpicture}[baseline,scale=0.5]
\begin{axis}[
ytick={-0.05,0.0,+0.05},
scale only axis,
hide x axis=true,
scaled y ticks = false,
y tick label style={/pgf/number format/fixed},
font=\LARGE,
title=\textbf{topic-comment}
]
\pgfplotstableread{plot-diff-topicxcomment-base-map}\table 
\addplot+[mark=+] table[x index=0,y index=1] from \table;
\addlegendentry{Web 2009}
\pgfplotstableread{plot-diff-2010-topicxcomment-base-map}\table 
\addplot+[mark=o] table[x index=0,y index=1] from \table;
\addlegendentry{Web 2010}
\addplot+[smooth] coordinates{(0,0) (50,0)};
\end{axis}
\end{tikzpicture} 
\end{tabular}
}
\caption{\label{fig:diff} Sorted per-query difference in MAP between the baseline and our model (y-axis), for each rhetorical relation. The horizontal line marks the baseline. $+$ and $o$ mark the 2009 and 2010 query sets.}
\end{figure*}

The scores shown in Table \ref{tab:scores} are averaged over tens of queries, meaning that they can be affected by outliers. Figure \ref{fig:diff} presents a detailed per-query overview of the performance of each query in relation to the baseline for each of the 15 rhetorical relations\footnote{Similar trends are observed in the corresponding figures for BPREF and NDCG, which are not included here for brevity.}. The plotted points represent the difference in MAP between our approach and the baseline. Positive points indicate that our approach outperforms the baseline. The points are sorted. 

 We observe that although the overall performance of the Web 2010 query set is lower than that of the Web 2009 query set, the improvements over the baseline of the 2010 set are consistently larger. Only in one case, \texttt{topic-comment}, do the plotted points clearly cross. Overall both query sets show similar plots with outliers at both ends of the scale. However, the 2009 query set tends to have a somewhat larger proportion of negative outliers, which goes some way towards explaining the lower improvements over the baseline observed for Web 2009. The Web 2010 set shows improvements over the baseline for most of the rhetorical relations and for the majority of the queries.

\subsubsection{Quantifying the Contribution of Rhetorical \\ Relations to the Ranking}
Exactly how much impact each rhetorical relation has on the ranking can be seen in Table \ref{tab:kappa}. The table lists the $\kappa$ values for the best performing tuned runs from Table \ref{tab:scores}, where high $\kappa$ values mean that the rhetorical relations are given more weight in the ranking (see Equation \ref{eq:mix}). We see that none of the values are above 0.5 for MAP and NDCG, indicating that too much emphasis on the rhetorical relations may not be beneficial to performance. Consistent with Table \ref{tab:scores}, BPREF follows a different trend than MAP and NDCG, which could be due to the fact that it is a different type of evaluation measure as discussed above in section \ref{sss:tuning}. With BPREF, unassessed documents are not explicitly penalised in the evaluation (as in MAP and NDCG) - resulting in overall higher $\kappa$ values for best performing runs, typically of around 0.5-0.7.

Further we observe that the rhetorical relations that consistently improve performance over the baseline, as indicated in Table \ref{tab:scores}, differ in $\kappa$ values for their best performing runs. For example, $\kappa$ = 0.2 - 0.3 for \texttt{background} and $\kappa$ = 0.5 for \texttt{topic-comment}. This implies that, to use rhetorical relations successfully for IR, it is not sufficient to know which rhetorical relations should be considered in the ranking and which not; also knowledge about how much emphasis to put on each rhetorical relation is needed for optimal IR performance.

Finally, note that the frequency of rhetorical relations does not affect their impact to retrieval. For instance, the three best performing rhetorical relations, \texttt{topic-comment, background} and \texttt{cause-result} constitute respectively approximately $>$1\%, 11\% and 5\% of all rhetorical relations, as shown in Figure \ref{fig:distr-rr}.

\section{Optimised Ranking with \\
Rhetorical Relations}
\label{s:Model2}
\subsection{Rhetorical Relation Selection}
\label{ss:Selection}
The findings in section \ref{ss:Results} show that some rhetorical relations can be more beneficial to retrieval performance than others. An ideal solution would not consider the lexical statistics of all rhetorical relations in a document, but rather it would select to include in the ranking only those rhetorical relations that have a higher likelihood of enhancing retrieval performance. This can be formulated as finding the optimal rhetorical relation $\hat{\psi}$ that maximises the expected retrieval scores according to an evaluation measure (e.g. MAP) for a query-document pair: 
\begin{equation}
\label{eq:argmax}
\hat{\psi}  = \arg\max_{\psi \in \Psi} E[y|q,d]
\end{equation}

\noindent where $E$ denotes the expectation and $y$ the retrieval score (rest of notation as defined in section \ref{s:Model}).

Bayesian decision theory allows to reason about this type of expectation, for instance see \cite{WangZ10}. In this work, we treat this as a problem of Bayesian posterior inference, where the goal is to estimate the retrieval performance associated with a rhetorical relation, given the observed retrieval scores it fetches on a number of queries. Then, we can consider the rhetorical relation associated with the highest retrieval performance as optimal. For this estimation, we split our dataset into different parts so that we use the observations from one to make inferences about the other (see section \ref{ss:OptimisedResults} for details). 

Let $n=15$ be the rhetorical relations shown in Table \ref{tab:scores}, and $x_j$ be the number of queries for which retrieval with the $j^{th}$ rhetorical relation gets a retrieval score $y_j$. For now we assume that all rhetorical relations may be expected to have similar retrieval performance, with the $j^{th}$ rhetorical relation having an average performance ratio per query $\lambda_j$ (estimated as $\frac{y_j}{x_j}$). Various densities can be used to fit similar data \cite{ManmathaRF01}, one of which is the Poisson distribution. Let us assume that, conditional on $\lambda_j$, the retrieval scores $y_j$ have independent Poisson distributions with means $\lambda_j x_j$.  Let us further assume that the $\lambda_j$ are independent realisations of a gamma variable with parameters $\alpha$ and $\beta$, and that $\beta$ itself has a prior gamma distribution with parameters $\nu$ and $\phi$. Thus
\begin{eqnarray*}
\label{eq:}
f(y|\lambda) = \prod^{n}_{j=1} \frac{(x_j \lambda_j)^{y_j}}{y_j!}e^{-x_j \lambda_j}\\
\pi(\lambda|\beta) = \prod^{n}_{j=1}\frac{\beta^{\alpha} \lambda^{\alpha - 1}_j}{\Gamma(\alpha)}e^{- \beta \lambda_j}\\
\pi(\beta) = \frac{\phi^{\nu} \beta^{\nu -1}}{\Gamma (\nu)} e^{- \phi \beta}
\end{eqnarray*} 

\noindent so that the joint probability density of the retrieval scores $y$, the average performance ratios $\lambda$, and $\beta$ is
\begin{equation}
\label{eq:density}
f(y| \lambda) f(\lambda | \beta) \pi(\beta) = c \prod^{n}_{j=1} \{ \lambda_j^{y_j + \alpha -1} e^{- \lambda_j (x_j + \beta)} \} \cdot \beta^{n \alpha + \nu - 1} e^{- \phi \beta}
\end{equation}
\noindent where $c$ is a constant of proportionality.

The conditional density of $\beta$ can be computed by various numerical approximations, one of which is the Laplace method \cite{AzevedoS:94}, which we use here. 
To find the conditional density of $\beta$ we integrate over the $\lambda_j$ to obtain
\begin{equation}
\label{eq:}
f(y,\beta) = c \prod^{n}_{j=1} \{ (x_j + \beta)^{-(y_j + \alpha)}  \Gamma(y_j + \alpha)\} \cdot \beta^{n \alpha + \nu - 1} e^{- \phi \beta}
\end{equation}

\noindent from which the marginal density of $y$ is obtained by further integration to give
\begin{equation}
\label{eq:}
f(y) = c \prod^{n}_{j=1} \Gamma(y_j + \alpha) \cdot \int^{\infty}_0 \! e^{-h(\beta)} d\beta
\end{equation}

\noindent where $h(\beta) = \phi \beta - (n \alpha + \nu - 1) log \beta + \sum (y_j + \alpha) log (x_j + \beta)$. Let $I$ denote the integral in this expression. In this work, we take an uninformative prior for $\beta$, with $\nu=0.1$ and $\phi=1$ and use $\alpha=1.8$\footnote{These values are not tuned; they are the default values of this approach as illustrated in \cite{Davison:2009}, chapter 11.3, pages 603-604.}.  We then apply Laplace's method to $I$, resulting in the approximate posterior density for $\beta$, $\tilde{\pi} (\beta | y) = \tilde{I}^{-1} e^{-h(\beta)}$.

To calculate approximate posterior densities for $\lambda_j$ we integrate Equation \ref{eq:density} over $\lambda_i$, $i \ne j$ and then we apply Laplace's method to the numerator and denominator integrals of 
\begin{equation*}
\pi (\lambda_j | y) = \frac{\lambda_j^{y_j+ \alpha-1} e^{- \lambda_j x_j} \int^{\infty}_{0}e^{-h_j(\beta)} d\lambda}{\Gamma(y_j + \alpha) \int^{\infty}_{0} e^{-h(\beta)} d\beta}
\end{equation*}

\noindent where

\begin{equation*}
h_j(\beta) = (\phi + \lambda_j) \beta - (n\alpha + \nu -1) log \beta + \sum_{i \ne j}(y_i + \alpha) log(x_i + \beta)
\end{equation*}

\noindent The resulting denominator is again $\tilde{I}_1$, while the numerator must be recalculated at each of a range of values for $\lambda_j$. 
The output is the (posterior) expected retrieval performance associated with each rhetorical relation.

\begin{table*}
\centering
\caption{\label{tab:opt-scores}Retrieval performance with optimal rhetorical relations (inferred, observed) and without rhetorical relations (baseline). (1)-(5) refers to the five randomised samplings used to infer the optimal rhetorical relations. Bold marks better than baseline. }
\scalebox{0.85}{
\begin{tabular}{|l||lr|lr|lr||lr|lr|lr|} 
\hline
\multirow{2}{*}{rhetorical relation} 
&\multicolumn{6}{c||}{Web 2009 (queries 1-50)} &\multicolumn{6}{c|}{Web 2010 (queries 51-100)}\\
    &\multicolumn{2}{c|}{MAP}    	&\multicolumn{2}{c|}{BPREF}  	&\multicolumn{2}{c||}{NDCG}  &\multicolumn{2}{c|}{MAP}    	&\multicolumn{2}{c|}{BPREF}  	&\multicolumn{2}{c|}{NDCG}  \\ 
\hline
none (baseline)             &0.1625&  	    &0.3230& 	        &0.3894& &0.0967&   &0.2198& &0.2890&\\
optimal$_{inferred}$ (1) 	&\bf0.1879&+15.6\% 	&\bf0.3503&+8.5\% 	    &\bf0.4224&+8.5\%  &\bf0.1355&+40.1\%   &\bf0.2859&+30.1\% &\bf0.3347&+15.8\%\\
optimal$_{inferred}$ (2) 	&\bf0.1948&+19.9\% 	&\bf0.3585&+11.0\% 	    &\bf0.4202&+7.9\%  &\bf0.1285&+32.9\%   &\bf0.2841&+29.3\% &\bf0.3394&+17.4\%\\
optimal$_{inferred}$ (3) 	&\bf0.1984&+22.1\% 	&\bf0.3532&+9.3\% 	&\bf0.4169&+7.1\%  &\bf0.1358&+40.0\%   &\bf0.2906&+32.2\% &\bf0.3388&+17.2\%\\
optimal$_{inferred}$ (4) 	&\bf0.1952&+20.1\% 	&\bf0.3479&+7.7\% 	&\bf0.4282&+10.0\%  &\bf0.1360&+40.6\%   &\bf0.2874&+30.8\% &\bf0.3336&+15.4\%\\
optimal$_{inferred}$ (5) 	&\bf0.1950&+20.0\% 	&\bf0.3528&+9.2\% 	&\bf0.4287&+10.1\%  &\bf0.1340&+38.6\%   &\bf0.2865&+30.3\% &\bf0.3322&+14.9\%\\
optimal$_{observed}$ 	    &\bf0.2157&+32.7\% 	&\bf0.3660&+13.3\%	    &\bf0.4412&+13.3\%  &\bf0.1474&+52.4\%   &\bf0.2978&+35.5\% &\bf0.3569&+23.5\%\\
\hline
\end{tabular}
}
\end{table*}

\subsection{Experiments}
\label{ss:OptimisedResults}

\subsubsection{Setup}
 The observations required to make the above inference are triples of \textit{rhetorical relation - query number - retrieval score}. To avoid overfitting, we pool randomly 50\% of the observations from the 2009 Web query scores and 50\% of the observations from the 2010 Web query scores. We use this pool to infer the expected retrieval performance of each rhetorical relation. We repeat this randomised pooling five times, each time randomly pertrubing the data, producing five different sets of observations. We then use each set to infer the expected best performing rhetorical relation per query, in accordance to Equation \ref{eq:argmax}. Following this, we use the model introduced in section \ref{s:Model}, Equation \ref{eq:2}, to rank documents with respect to queries only for optimal (as inferred) rhetorical relations. We evaluate the above method using the same experimental settings described in section \ref{ss:Settings}. 

\subsubsection{Findings}
Table \ref{tab:opt-scores} shows the runs corresponding to the five different inferences of the best rhetorical relation that use our model (optimal$_{inferred}$ (1)-(5) respectively). We also report the optimal retrieval performance actually observed in the dataset when using the best rhetorical relation per query (optimal$_{observed}$). Optimal here means with respect to the choice of rhetorical relation, not with respect to the Dirichlet $\mu$ parameter of the baseline retrieval model. 

Table \ref{tab:opt-scores} shows that our optimised ranking model for rhetorical relations is better than the baseline for any of the five random inferences on all three evaluation measures. The probability of getting such a positive result by chance is $\frac{1}{2^5} < 0.05$, and thus the improvements are statistically significant. The improvements over the baseline are considerable, a very promising finding given the relatively low number of observations used for optimising the choice of rhetorical relations. Experiments involving larger query sets can be reasonably expected to perform on a par with state-of-the-art performance. 

More generally, the improvements in Table \ref{tab:opt-scores} signal that rhetorical relations (derived automatically as shown in this work) could potentially be useful features for `linguistically-uninformed' learning-to-rank approaches.

\section{Discussion}
\label{s:Discussion}

\subsection{Rhetorical Relation Distribution}
The distribution of the 15 rhetorical relations we identified in our dataset is not the same for all rhetorical relations (see Figure \ref{fig:distr-rr}). Some types, e.g. \texttt{topic-comment}, tend to be very sparse, whereas relations such as \texttt{elaboration} prevail. This has no impact on the model presented in section \ref{s:Model}, but it can bias the optimised inference of the model presented in section \ref{s:Model2}. The lower the occurrence of a rhetorical relation in the dataset, the fewer the observations of retrieval performance associated with it, and hence the weaker the predictions we can infer about whether it is optimal or not. A fairer setting would be to have the same number of `query - retrieval performance' observations for all rhetorical relations - however that would imply fiddling with the document distribution of our dataset significantly, potentially harming its quality as a test collection.

\subsection{Limitations}
A general limitation of discourse analysis is that not all types of text are susceptible to it. For instance, legal text, contracts, or item lists often lack rhetorical structure. In this work, we made no effort to identify and exempt such types of text from the discourse parsing. We reasoned that, as the SPADE parser includes a first-step grammatical parsing, the initial grammatical parsing of these types of text would flag out ill-formed parts (e.g. missing a verb, or consisting of extremely long sentences), which would then be skipped by the discourse analysis. This was indeed the case, however at a certain efficiency cost. Overall processing speed for SPADE was approximately 19 seconds per document (including the initial grammatical parsing), on a machine of 9 GB RAM, 8 core processor at 2.27GHz. One way of improving this performance would be to update the first-step grammatical parsing. Currently this depends on the well-known Charniak parser \cite{Charniak:2000:NAACL}, which is one of the best performing grammatical parsers, however no longer supported. Other state-of-the-art faster grammatical parsers, e.g. the Stanford parser\footnote{http://nlp.stanford.edu/software/lex-parser.shtml}, could be adapted and plugged into SPADE instead. 

The choice of applying out model for re-ranking as opposed to ranking all documents was closely related to the efficiency concerns discussed above. Our model is not specific to re-ranking only, however, using SPADE on more than 50 million documents was too expensive at this point. Improving the discourse parser's efficiency is something we are currently working on, with the aim to apply our model for full ranking and see if the conclusions drawn from this work hold.

Finally, the accuracy of the discourse parser was not considered in this work, apart from indications in the literature that SPADE is a generally well-performing parser \cite{SoricutM03}. Given that the default version of the parser we used is trained on news articles, one may reason that its accuracy could improve if we train it on the retrieval collection, or on documents of the same domain. Note that, parsing accuracy aside, rhetorical relations assignment is not an entirely unambiguous process, even to humans \cite{MannT:1988}. For the purposes of this work, this type of fine-grained ambiguity may however not be important to retrieval performance.

\subsection{Future Extensions}
\label{ss:future}
Future extensions include primarily making SPADE scalable on large collections of documents as discussed above, as well as using more than one rhetorical relation per document. For instance, the posterior probabilities estimated in section \ref{ss:Selection} could be used to weight the text in each rhetorical relation. If those posteriors are too flat, an exponent could make them peakier. As the exponent goes to infinity, the maximum relation model presented in section \ref{ss:OptimisedResults} would be recovered. In addition, we intend to refine the discourse analysis by considering the nucleus (i.e. central) versus satellite (i.e. peripheral) rhetorical relations for IR, as well as to improve the effectiveness of the discourse parser by training it on data of the same domain. As discussed in section \ref{ss:Induction}, we will also investigate alternative estimations of Equations \ref{eq:2}-\ref{eq:mix}.

 An interesting future research direction is the potential relation between rhetorical relations and user context: for instance, in a search session including several query reformulations, is there a correlation between the progression of the information need of the user and the rhetorical relations that the retrieval system should boost in a document (e.g. \texttt {elaboration}), as indicated by Sun \& Chai \cite{SunC:2007}? Another interesting future extension of this work is in relation to evaluation measures of graded relevance measures on an inter-document level, as investigated in XML retrieval \cite{2009Lalmas} for instance. If parts of a document can be regarded as more or less relevant, this may be reflected to their discourse structure. This might be especially useful for multi-threaded documents, such as multiple-user reviews and opinions, where the discourse relations tend to shift markedly. Finally, the current operationalisation of our model is simplistic in the sense that the term `rhetorical relation' is coerced into meaning `non-overlapping text fragment' and the actual relation between bits of text is discarded in the process. In future work we could apply fielded XML retrieval models in order to investigate nested structuring among rhetorical relations.

\section{Conclusions}
\label{s:Conclusions}
Rhetorical relations, e.g. \texttt{contrast, explanation, condition}, indicate the different ways in which the parts of a text are linked to each other to form a coherent whole. This work studied two questions: Is there a correlation between certain rhetorical relations and retrieval performance? Can knowledge about a document's rhetorical relations be useful to IR? To address these, we presented a retrieval model that conditions the probability of relevance between a query and a document on the rhetorical relations occurring in that document. We applied that model to an IR re-ranking scenario for Web search. Experimental evaluation of different versions of our model on TREC data and standard settings demonstrated that certain rhetorical relations can be beneficial to retrieval, with $>$10\% improvements to retrieval precision. Furthermore, we showed that these improvements over the baseline can improve significantly, when the optimal rhetorical relation per document is selected for retrieval. 

Overall, three rhetorical relations were found to benefit retrieval performance notably and consistently for different evaluation measures and query sets: \texttt{background, cause-result} and \texttt{topic-comment}. In retrospect, this is perhaps not surprising, since these are among the most salient discourse relations on an intuitive basis: the main topic or theme of a text, its background, causes and results \cite{LouisJN10}. Future extensions and research directions of this work include applying our model for ranking all documents (as opposed to re-ranking only) and experimenting with alternative estimations of its components.




\section{Acknowledgments} We thank Kasper Hornb\ae k, Jakob Grue Simonsen, Raf Guns, Qikai Cheng and the anonymous reviewers for helping improve this paper. Work partially funded by the Danish International Development Agency DANIDA (grant no. 10-087721) and the National Natural Science Foundation of China (grant no. 71173164).
%

%
%
\balancecolumns
\end{document}